# Global Health Monitor - A Web-based System for Detecting and Mapping Infectious Diseases

Son Doan*, QuocHung-Ngo¶, Ai Kawazoe*, Nigel Collier*

\* National Institute of Informatics,
2-1-2 Hitotsubashi, Chiyoda-ku, Tokyo,
Japan
{doan,zoeai,collier}@nii.ac.jp

¶ University of Information Technology,
Vietnam National University (HCM),
Vietnam
hungnq@uit.edu.vn

**Abstract**

We present the Global Health Monitor, an online Web-based system for detecting and mapping infectious disease outbreaks that appear in news stories. The system analyzes English news stories from news feed providers, classifies them for topical relevance and plots them onto a Google map using geo-coding information, helping public health workers to monitor the spread of diseases in a geo-temporal context. The background knowledge for the system is contained in the BioCaster ontology (BCO) (Collier et al., 2007a) which includes both information on infectious diseases as well as geographical locations with their latitudes/longitudes. The system consists of four main stages: topic classification, named entity recognition (NER), disease/location detection and visualization. Evaluation of the system shows that it achieved high accuracy on a gold standard corpus. The system is now in practical use. Running on a cluster-computer, it monitors more than 1500 news feeds 24/7, updating the map every hour.

## 1 Introduction

Information concerning disease outbreak events is published in various news outlets on the World Wide Web, in many different languages. Identifying early news stories about disease outbreaks *automatically* is important for a bio-surveillance system that is designed to inform health professionals. Currently, there are several systems available for the disease detection and tracking task. For example, ProMED-mail (2001) or MedISys (2007) (Medical Intelligence System). ProMED-mail is an Internet-based system that provides reports by public health experts concerning outbreak diseases (that is, the system is *not* automatic but rather human curated). In contrast to ProMED-mail, MediSys is an automatic system working on multilingual languages, but it mainly focuses on analyzing news stories based on the country level. Another system which is close to the one we present is HealthMap (Brownstein and Freifeld, 2007). HealthMap automatically collects news from the Internet about human and animal health and plots the data on a Google Maps mashup. Data is aggregated by disease and location. Unlike HealthMap, our system takes an ontology-centred approach to knowledge understanding and linkage to external resources. For annotation of topics and entities we also exploit a range of linguistic resources within a machine learning framework.

There are several challenges in geo-coding when dealing with news stories. The two main challenges are disease/location extraction and geo-disambiguation. The former is concerned with how to determine disease and location names for disease-related news stories. The latter is concerned with how to solve geo-disambiguation. For example, if there is a news story about equine influenza in Camden, the system should detect that the disease name is "equine influenza" and the location name is "Camden". However, there are two locations named Camden: One in Australia and one in

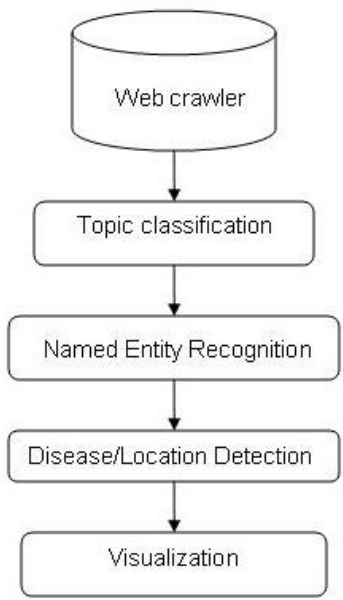

**Figure 1. Stages of the system.**

London, UK. The problem is that only one location should be chosen for plotting into a map. In our opinion, current systems lack the advanced natural language processing and text mining techniques that would enable the automatic extraction of such disease/location event information.

BioCaster is a project working towards the detection and tracking of infectious diseases using text mining techniques. One of the main components is the BioCaster Ontology (BCO), which includes 50 infectious diseases with links to external vocabulary resources, and a geographical ontology of a total of 243 countries and 4025 sub-countries (province and cities) with their latitudes/longitudes. We also now automatically link news on outbreaks to academic sources such as Stanford university's Highwire and NCBI's PubMed using search terms Disease name + Location name (Country) + "case". This is to ensure a focus on recent case report relevant to the news items.

The system includes four main stages: topic classification, named entity recognition (NER), disease/location detection, and visualization. The current version of the system (English only) can be found at http://biocaster.nii.ac.jp.

The remainder of this paper is organized as follows. Section 2 outlines the BioCaster Ontology (BCO). Section 3 describes some features of the system (modules, functionality and algorithms). Section 4 is concerned with system evaluation. Finally, Section 5 outlines conclusions and presents possible future work.

## 2 Overview of BCO

BCO is one of the main components in the BioCaster project. It includes an ontology of 50 infectious diseases and a geographical ontology (243 countries and 4,025 sub-countries). The infectious disease ontology was built by a team consisting of a linguist, an epidemiologist, a geneticist, and a medical anthropologist. A disease in BCO has a root name which is unique identifier and also other properties relating to synonyms, symptoms, associated syndromes, hosts, etc. The ontology is multilingual, supporting six languages (English, Japanese, Vietnamese, Thai, Korean, and Chinese); and has links to external ontologies (such as MeSH, SNOMED and ICD9/10) and resources (like Wikipedia). The geographical part is built from Wikipedia[1]. The BCO is available on the Web at http://biocaster.nii.ac.jp. For a fuller description of the BCO, see Collier et al. (2007a) and Kawazoe et al. (2007).

## 3 The System

### 3.1 Overview of the system

The Global Health Monitor system runs on a cluster machine with 16 nodes under the Linux operating system. The code was written in PHP, Perl, C, and Java and the number of input news feeds is about 1,500. The system has a crawler that collects news every hour. These collected news stories are then processed and analyzed step-by-step in four main phases: topic classification, named entity recognition (NER), disease/location detection, and visualization. Each of the four phases is managed by a distinct module. These components are depicted in Figure 1. The first three modules are run inside the system and the visualization module – the Google Map – can be seen at the BioCaster portal. Figure 2 shows a screenshot of the system.

---

[1] http://www.wikipedia.org.

952

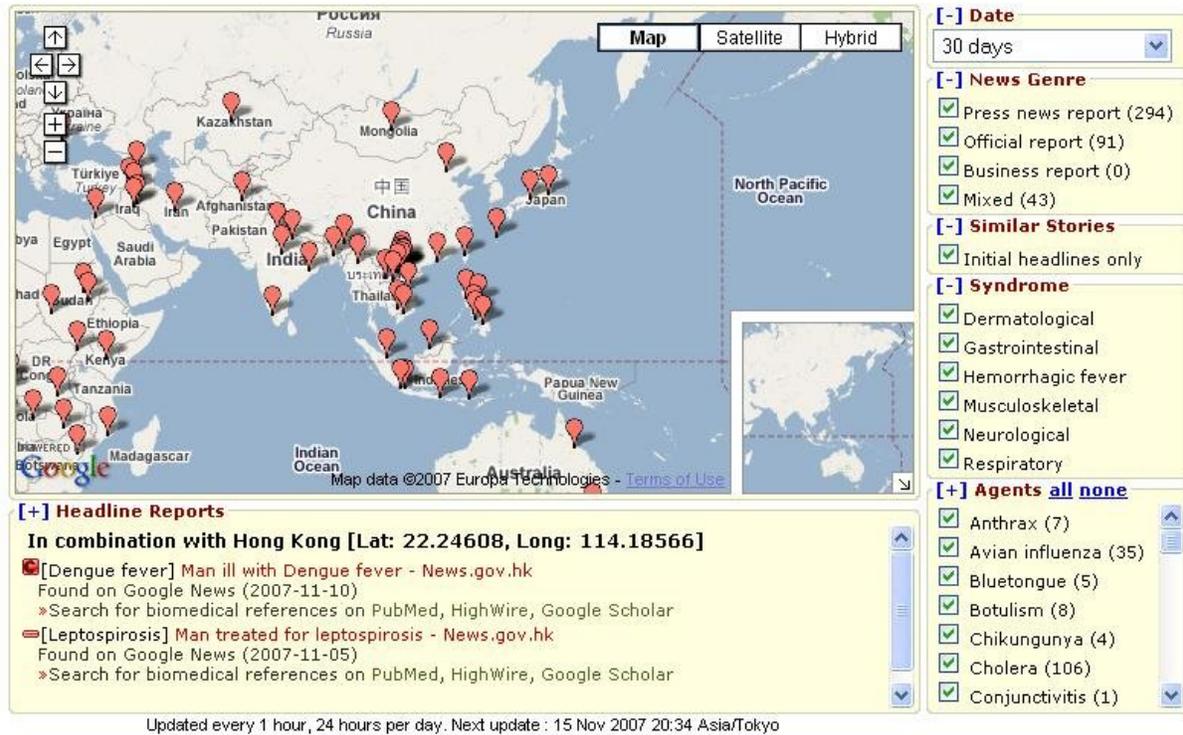

**Figure 2. The Global Health Monitor system, showing disease events from the last 30 days. The main screen is a Google Map. Selected headline reports run along the bottom of the screen and link to biomedical reference on PubMed, HighWire and Google Scholar. Symbol**  **links to disease names in the BCO and symbol**  **stands for disease name not in the BCO. The right of the screen shows various user options to filter the news.**

We will now describe each of the four modules in turn:
**\* Topic classification.** This module identifies news stories with disease-related topics and retains relevant ones for later processing. The module uses ontology-supported text classification with naïve Bayes as the classification algorithm and the Bio-Caster gold standard corpus as the training data set (Doan et al., 2007). In this module, we used the Rainbow toolkit.[2]

**\* NER.** Disease-related news stories are automatically analyzed and tagged with NEs like PERSON, ORGANIZATION, DISEASE, LOCATION. This module is implemented by SVM classification algorithm[3]. For a more detailed description of the schema and NER module, see Kawazoe et al. (2006).

**\* Disease/location detection.** This module extracts disease and location information. Details are given in Section 3.2.

**\* Visualization.** The detected locations are plotted onto a Google map with ontology links to associated diseases and news stories.

### 3.2 Disease/location detection algorithm

The disease/location detection algorithm is based on a statistical model of the LOCATION and DISEASE Named Entities (NEs). The algorithm can be described as follows:

---

[2] Rainbow toolkit, available at http://www.cs.umass.edu/~mccallum/bow/rainbow

[3] TinySVM, available at http://chasen.org/~taku/software/TinySVM.

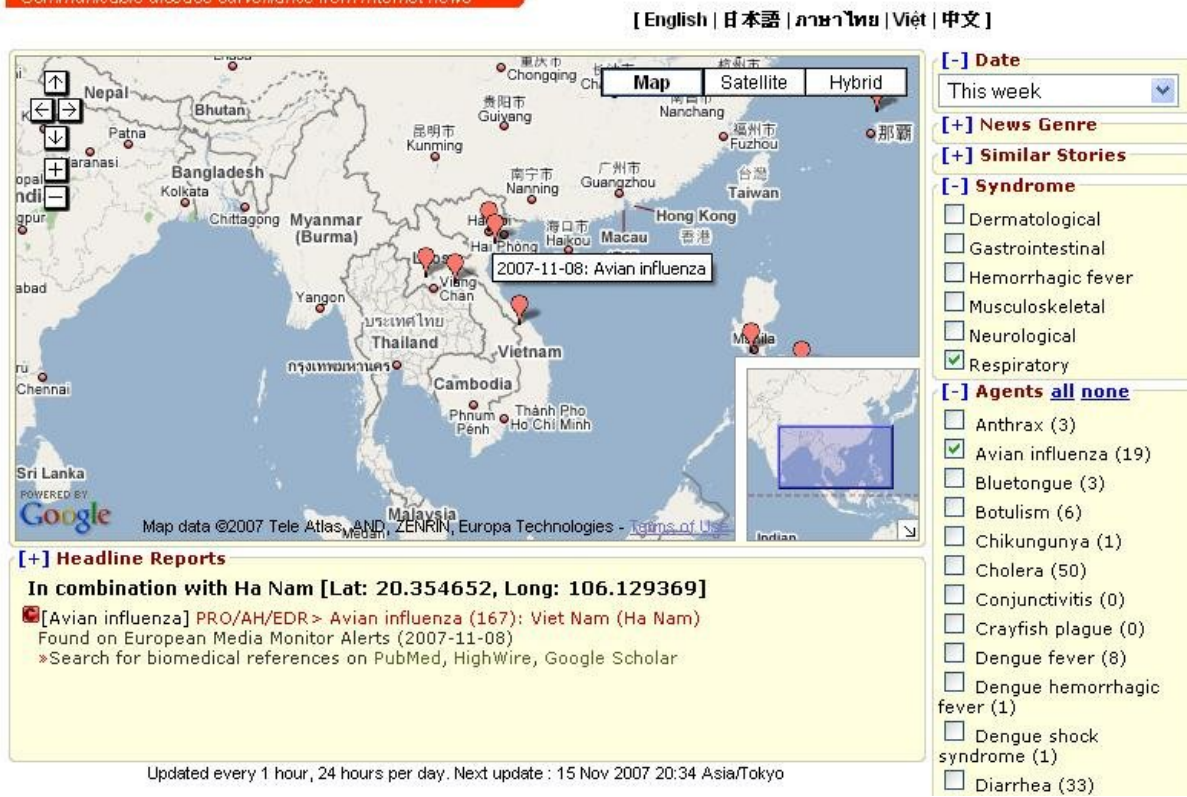

**Figure 3. The Global Health Monitor with the *Respiratory Syndrome* selected. The time span selected is the current week.**

**Input:** A set of news stories tagged with NEs.
**Output:** A set of disease/location pairs.

**Step 1:** Detect LOCATION-DISEASE pairs in each news story by corresponding NEs, and calculate their frequency in a news story.

**Step 2:** Calculate the frequency of LOCATION-DISEASE pairs in a corpus.

**Step 3:** Rank LOCATION – DISEASE pairs by the frequencies calculated in Step 2. Use a threshold to choose top LOCATION - DISEASE names[4].

**Step 4:** Map disease and location names: If DISEASE matches to a synonym in BCO then DISEASE was assigned to that disease name. This process of matching (grounding the terms) allows us to provide extra information from the ontology and to remove variant forms of terms from the map – thereby aiding readability. Similarly, if LOCATION matches to a location in BCO then LOCATION was assigned to that location name.

**Step 5:** Re-map into news stories: Match detected diseases and locations within the first half of each news story. If both disease and location are matched then they are stored; otherwise, skip.

This five step process is repeated every hour, for each news article that is less than 1 day (24 hours) old.

### 3.3 Capabilities of the system

The following lists some capabilities of the current Global Health Monitor system.

\* Date range: The system shows the disease/location and news stories within a specific date range. Current implemented date ranges are: 30 days ago, 3 weeks ago, 2 weeks ago, 1 week ago, this week and today.

---
[4] In the current system, we set the threshold value to 40.

* Genre filter: The system can show news stories by publication type. There are four genres of news: Press news (like Google News, Yahoo News), Official news (like ProMED, WHO reports), Business news, and Mixed news (like Topix.com).

* Similar stories: The system currently uses a simple method to remove duplicate news stories. Users can use the "Initial headline only" option to activate this function.

* Syndrome filter: There are six syndromes in BCO: Dermatological, Gastrointestinal, Hemorharrgic fever, Musculoskeletal, Neurological, and Respiratory. A syndrome can be associated with several diseases included in BCO. The system can show news stories related to these syndromes.

* Agent option: This option allows users to view lists of infectious diseases which come from BCO. Some diseases though are not in the BCO. Users can choose some, all, or no diseases using a checkbox style interface.

Figure 3 shows the interface when users choose Syndromes as Respiratory for this week at the current view.

## 4 Evaluation

To evaluate any bio-surveillance system is very challenging (Morse, 2007). Our system is an integration of several modules, e.g., classification, NER and other algorithms. The evaluation processes for these modules are briefly described below:

### 4.1 Topic classification

Evaluation of topic classification is presented in Doan et al. (2007). The system used the BioCaster gold standard corpus which includes 1,000 annotated news stories as training data. The classification model is naïve Bayes with features as raw text, NEs, and Roles (Doan et al., 2007). The system achieved an accuracy score of 88.10%.

### 4.2 NER evaluation

The evaluation of the NER system module is reported in Kawazoe et al. (2006). We used an annotated corpus of 200 corpus news articles as training data. The NER system achieved an F-score of 76.97% for all NE classes.

### 4.3 Disease/location detection

For the preliminary evaluation of disease/location detection, we used data from a one-month period (from October 12 to November 11, 2007).

In our observations, the system detects about 25-30 locations a day, an average of 40 infectious diseases and 950 detected pairs of diseases/locations per month (A news story can contain multiple locations and diseases). The main news resources mostly come from Google News (251 pairs, about 26.4%), Yahoo News (288 pairs, about 30.3%), ProMED-mail (180 pairs, about 18.9%), and the remaining 24.3% for others. The running time for updating disease/location takes about 5 minutes.

In order to evaluate the performance of disease/location detection, we define the Precision and Recall as follows:

$$\text{Precision} = \frac{\#\text{Relevant pairs} \cap \#\text{Retrieved pairs}}{\#\text{Retrieved pairs}}$$

$$\text{Recall} = \frac{\#\text{Relevant pairs} \cap \#\text{Retrieved pairs}}{\#\text{Relevant pairs}},$$

Where #Relevant pairs is the number of disease/location pairs that human found, and #Retrieved pairs is the number of disease/location pairs that the system detected.

The Precision can be calculated based on our retrieved pairs detected by the system, however the Recall is under estimated as it does not measure pairs missed by the system in the topic classification stage.

We evaluate the Precision of disease/location detection on 950 pairs of location/disease. The system correctly detected 887 pairs, taking 887/950=93.4% Precision.

### 4.4 Limitations of the system

There are some limitations of the system. The first limitation is there are several cases of ambiguity. For example, news stories about "A team at Peking University in Beijing studied tissue taken from 2 people killed by H5N1 in China" or "A meeting on foot and mouth disease (FMD) was held in Brussels on 17<sup>th</sup> October, 2007". The system incorrectly detects the location as Beijing in the first story, and Brussels in the second one. Another hard case is location disambiguation, e.g., news about "Rabies in Isle of Wight" in which in the main body does not mention anything about country and sub-country. There are two locations named "Isle of Wight" in our geo-ontology: one in Virginia, USA and one in the UK. In the future, we will look at the country-level information of new providers (by checking domain names) to solve this problem. For example, if a news story mentions the Isle of Wight, and the news story originates from the UK, then it will be taken to refer to the Isle of Wight in the UK.

The second limitation is the ability to detect new diseases or locations that are not in the ontology. In the future work, we will augment newly detected diseases as well as improve the geographical ontology.

## 5 Conclusion

We presented the Global Health Monitor - a Web-based system for detecting and mapping infectious diseases from Web. The system collects news from news feed providers, analyzes news and plots disease relevant data onto a Google map. Preliminary evaluations show that our system works efficiently with real data.

In the future, we will develop more efficient algorithms for detecting diseases/locations based on relation identification. Named relation will be described in the BCO event taxonomy (Kawazoe et al., 2007). Extra capabilities will be added to the system like classifying outbreak of disease by countries, detecting new diseases that are not in out ontology, and showing timeline of news stories. Evaluation of the timeliness system against human curated sources like ProMED-mail will be implemented. Working versions for other languages like Vietnamese, Japanese, and Thai are also being considered, using the existing BioCaster disease ontology.

## Acknowledgements

The authors wish to thank Mike Conway at the National Institute of Informatics for revising the manuscript, and both Mika Shigematsu and Kiyosu Taniguchi at the National Institute of Infectious Diseases for useful discussions. This work was supported by Grants-in-Aid from the Japan Society for the Promotion of Science (grant no. 18049071).